\documentstyle[12pt]{article}

\title{WHAT ARE THE BUILDING BLOCKS OF OUR UNIVERSE?}

\author {Kameshwar C. Wali \\
Physics Department, Syracuse University \\
Syracuse, NY 13244-1130. U.S.A. }

\begin{document}
\maketitle
\begin{abstract}

We are told that we are living in a Golden Age of Astronomy. Cosmological Parameters are
found with un precedented accuracy. Yet, the known form of matter forms only a small
fraction of the total energy density of the universe. Also, a mysterious dark energy
dominates the universe and causes acceleration in the rate of expansion.

\end{abstract}

\section{Introductory Remarks}

We live in an exciting age of astronomy. Some thirty years ago, cosmology was a science
of only two parameters, the current expansion rate or the Hubble constant,$H_0$, and its
change over time or the deceleration parameter, $q_0$,. Questions such as the age of the
universe, its large and small scale structure, origin of galaxies and the formation of
stars were considered as speculative with no direct connection to precise measurements.
Situation has changed drastically with the discoveries of giant walls of galaxies,
voids, dark matter on the one hand,and on the other hand, the tiny variations in the
cosmic background radiation and a 'mysterious' uniformly distributed, diffuse dark
energy causing acceleration of the expansion rate of the universe. There are some
sixteen cosmological parameters whose measured values exhibit unprecedented accuracy in
the history of astronomy. Ten of these parameters are "Global" in the sense that they
pertain to the idealized standard model of a homogeneous isotropic universe governed by
Friedman-Laimetre-Walker-Robertson metric within the framework of general relativity.
The other six refer to more details of the model, to the deviations from homogeneity and
their manifestations in the cosmic structure.These numbers are tied to a fundamental
theory--big bang, inflationary theory and is believed by the practitioners that it
accounts for \emph{\textbf{the origin of structure and geometry of the universe, as well
as describing its evolution from a fraction of a second}}.

In the words of Freedman and Turner\cite{FT},the still evolving and emerging picture is
described as follows:

 \emph{In a tiny fraction of a
second during the early history of the universe, there was an enormous explosion called
inflation.This expansion smoothed out wrinkles and curvature in the fabric of
space-time, and stretched quantum fluctuations on subatomic scales to astrophysical
scales.Following inflation was a phase when the universe was a hot thermal mixture of
elementary particles, out of which arose \textbf{all the forms of matter that exist
today.}Some 10,000 years into its evolution, gravity began to  grow the tiny lumpiness
in the matter distribution arising from quantum fluctuations into the rich cosmic
structures seen today, from individual galaxies to the great clusters of galaxies and
superclusters}

However, there are wrinkles and surprises in this rosy theoretical picture!\cite{T}.
Most of the universe is made of some thing fundamentally different from the ordinary
matter that we know of. Some 30 percent of the total-mass energy density is \textbf{DARK
MATTER}, whose nature we do not know, but in all likely hood, they are composed of
particles formed in the early universe. About 66 percent is in the form of a smooth,
uniformly diffused energy called the \textbf{DARK ENERGY}, whose nature we do not
know,but we conjecture that its gravitational effects are responsible for the recently
observed acceleration in the rate of expansion of the universe\cite{evidence}.
Approximately only 4 percent is composed of ordinary matter, the bulk of which is dark.
Finally,cosmic microwave background radiation contributes only 0.01 percent of the
total, but it encodes \textbf{information about the space-time structure of the
universe, its early history, and probably even about its ultimate fate}.

In light of this, one wonders whether our present fundamental theories of elementary
particles that are supposed to be the building blocks of the universe are of any
relevance to the emerging picture of the universe. In this review, I present certain
aspects concerning the current status of particle theory and its link to cosmology.

\section{Beyond The Standard Model; Grand Unified Theories}

The current theory of fundamental interactions is the so called Standard Model, a
non-Abelian Yang-Mills type theory based on the gauge group $S(U(3)\times U(2))$ with
spontaneous symmetry breaking, induced by a fundamental scalar, called the Higgs meson.
It presents a unified theory of weak and electromagnetic interactions(electro/weak)
marked by spontaneous symmetry breaking. Strong interactions are described by the gauge
theory based on the group $SU(3)$ (Quantum Chromodynamics). It has been enormously
successful in its confrontation with experiments. Yet, it is far from a fundamental
theory for a number of reasons. First and foremost, it has a large number of free
parameters. The starting point is three families of quarks and leptons with their masses
totally arbitrary ranging over several orders of magnitude. The theory is
renormalizable, but it has quadratic divergences requiring "fine tuning" of the
parameters in successive orders of perturbation. It can accommodate CP violation, but
has no natural explanation for its origin or the order of magnitude of its violation.

Nonetheless, its enormous success led to its natural extension seeking unification of
all the three fundamental interactions, weak, electromagnetic and strong:\emph{Grand
Unified Theories (GUTS).} In its most pristine form, a grand unified theory postulates
that the description of interactions among elementary particles will simplify enormously
at some very high energy $E > M_G$ (Grand Unification Mass).The electro/weak and strong
interactions, which are the basic interactions at low or present laboratory energies,
will be seen as different aspects of one basic interaction among \emph{a set of basic
constituents of all matter.} Correspondingly, as one moves up in energy, a symmetry
larger than the standard model gauge group $S(U(3)\times U(2))$ will progressively
unfold itself, becoming fully manifest at energies exceeding $M_G$. Initial analysis
based on renormalization group methods suggested strongly that the coupling constants
that change as a function of energy(a feature of non-abelian gauge theories) evolve to a
unification point at energies around $10^{15} GeV$. Since any such unification demanded
quarks and leptons to be treated on the same footing, quark-lepton transitions at such
energies and above became theoretically mandatory, leading to the possible violation of
the well established baryon- and lepton- number conservation laws at low energies. A
dramatic consequence was the possibility of observing proton decay! The simplest
extension of the standard model based on the gauge group $SU(5)$ predicted a life time
of $10^{29}$ years for the proton and led to a number of experiments that failed to
detect it and have set an a limit to proton life time beyond $10^{32}$ years. More
complicated models based on bigger simple groups ($SO(10)$, for instance), semi-simple
product of groups, and exceptional groups (such as $E_6$) were proposed and were
partially successful in extending the predicted lifetime of the proton and predicting
new exotic species of particles.

However, to obtain a full display of the new interactions and to put them to
experimental test, we need energies of the order of $10^{15}GeV$ and greater, which are
clearly beyond the present or future terrestrial accelerators. It became evident that
astrophysics and cosmology were the natural arena for testing these ideas. In the
current popular standard model cosmology, based on Friedman-Laimetre-Walker-Robertson
metric, the early universe was in a hot dense phase with temperatures exceeding $10^{16}
GeV$ in its first $10^{-35}$ seconds after the big bang. The universe in its early
stages was like a giant accelerator and one expected a copious production of all the
particles we know, and those those we do not know - the super heavy particles predicted
by grand unified theories. One could then trace the effects of the new particles and
their interactions through the subsequent adiabatic cooling of the universe down to
present epoch and compare them with astrophysical measurements. It was the beginning of
a symbiotic relation between particle physics and astrophysics.

\section{Beyond The Standard Model; Supersymmetry}

Supersymmetry goes beyond the conventional distinction between fermions (odd integral
multiple of spin 1/2 particles) as fundamental constituents of matter and bosons (
integral multiples of spin 1 particles)as carriers of interactions. It treats both on an
equal footing, combining them in a supermultiplet that allows symmetry transformations
between them. Conventional space-time symmetries are supplemented by anti-commuting
operators that transform fermion into boson and vice versa. Thus it may be looked upon
as unification of matter and interactions.\\

Its main points are: \begin{itemize}\item {Each chiral fermion (quark,lepton) in the
standard model is accompanied by a spin zero boson (squark, slepton). Likewise each
gauge boson and Higgs scalar is accompanied by a spin 1/2 fermion (gaugino, Higgsino)
}\item{All superpartners of standard model(SM) particles are \emph{new
particles}}\item{No known SM particle is a superpartner of another SM particle. If
supersymmetry were exact, particle and its superpartner that have the same quantum
numbers should be degenerate in mass}\item{Supersymmetry is an \textbf{approximate}
symmetry of nature. If it were exact,superpartners of SM particles would have been
discovered along with the SM particles since they would have been degenerate in mass }
\end{itemize}

 From theoretical point of view, supersymmetry is very appealing. It
 is a beautiful symmetry, but it is approximate. There is no unique
 or elegant symmetry breaking mechanism. In principle, it has the
 potential of
 solving some theoretical problems associated with the quadratic
 divergences and fine tuning problems generic to the standard model
 and grand unified theories, which invoke spontaneous symmetry breaking
 through fundamental scalar particles. There is enough freedom in models
 to meet the experimental limits on proton life time exceeding
 $10^{32}$ years. The so called Minimal
 Supersymmetric Standard Model (MSSM), an extension of the standard
 model, provides a more convincing evidence for the unification of all
 interactions (excluding gravity)than grand unified theories alone. From the
 point of view cosmology and astrophysics,
 broken supersymmetry,offers a candidate for dark matter, the "neutralino."\cite{Feng}

 \section{Nature of Dark Matter; Candidates for Dark Matter}

 Observationally,dark matter appears to be distributed diffusively in
 external halos
 around individual galaxies or in a sea through which galaxies move.
 Here are some speculations concerning its
 nature:
 \begin{itemize}\item{It is believed to consist of hypothetical
 particles called WIMPS (Weakly Interacting Massive Particles),produced probably
 in the early universe}

\item {Their masses should be around electro/weak symmetry breaking scale, in the $10
GeV - 1 TeV$ range. They should have \textbf{neither strong or electromagnetic}
interactions with the known SM particles. If they did, the argument goes, they would
have dissipitaed energy and relaxed to more concentrated structures, where only known
baryons are found.} \item  {They must be \textbf{Cold}, in the sense that they move
slowly with non-relativistic velocities, as opposed to \textbf{hot}light particles
moving with relativistic velocities. \textbf{Hot} and \textbf{Cold} dark matter lead to
different predictions regarding galaxy formation. Galaxies are formed first due to cold
dark matter before forming superclusters, whereas opposite is what happens with hot dark
matter.}
 \end{itemize}

It is remarkable that from the simple starting point of cold dark matter and
inflation-induced lumpiness,one can envisage a highly successful picture of formation of
structure in the universe. From the point of view of particle physics, there are three
possible candidates for dark matter:
\begin{itemize}\item{\textbf{Neutrinos}: The idea that neutrinos could be candidates for dark matter
has been there for a long time. They certainly exist in large numbers (roughly one
billion for every photon) and they could contribute a huge mass to the dark matter if
they were massive enough. Recent experiments on solar and atmospheric neutrino
oscillations have established that one or more than one of the neutrinos must have a
mass. However, neutrino oscillation experiments probe only the mass differences.
Consequently, there are a number of theoretical models and experiments to determine
their absolute masses. Cosmological observations will play a very important role in
setting the absolute scale of neutrino mass just as primordial nucleo-synthesis set a
limit on the number of light neutrinos. This is because, as mentioned above, hot and
cold dark matter predict entirely different course for the evolution of the large scale
structure. If all the neutrinos are light with masses of an electron volt or less, they
constitute hot dark matter. Then, there is a stringent limit on the amount of hot dark
matter in order that it does not wipe away the required small scale structure.
}\item{\textbf{Axions}: Axion is probably the first candidate for dark matter that was
proposed. Its search has been going on for quite some time.It has its origin in the
theoretical solution of
 CP violation in strong interactions due to the complex nature of the vacuum in the
 of theory of strong interactions based on quantum chromodynamics (QCD). A global-axial
 symmetry known as Pecci-Quinn symmetry solved the problem, but it made it necessary to
 have a massive particle with strong interactions with ordinary matter. When
 experiments failed to detect such particle, a mechanism proposed by Dine, Fisher
 and Schrednicki, allowed the coupling to matter as well as its mass
 arbitrarily small. \emph{Axion exists, but it cannot be seen}.\\

 Two different mechanisms have been proposed for their production in the early universe
 (a). at the QCD phase transition, when free quarks get bound to form hadrons, a Bose
 condensate of axions form and these very cold particles behave as cold dark matter (b).
 Decay of cosmic strings at the Pecci-Quinn phase transition can also give rise to
 axions.\\

 Axions are potentially detectable through their weak couplings to electromagnetism. In
 the presence of a strong magnetic field, the axionic dark matter can decay into two
 photons. Several new experiments based on cryogenically cooled cavity and the use of an
 atomic beam of Rydberg atoms as a detector are in progress.}
 \item {\textbf{Neutralinos}: Broken supersymmetry combined with the conservation of what is
  called R-Parity provides an ideal
 candidate for dark matter. The lightest particle is absolutely stable and has the
 necessary properties to form dark matter. In MSSM, the spin 1/2 neutral gauge
 eigenstates, mix and form mass eigenstates after symmetry breaking. These are called
 \emph{Neitralinos}. The lightest among these is considered to be the most probable
 candidate for dark matter.\\

 Neutralinos are Majorana particles. Their mass estimates in MSSM depends upon five
 parameters. In order to estimate their contribution to relic dark matter density, it is
 necessary to know their annihilation cross-sections into ordinary as well as the
 superpartners. Such calculations have been made and restrictions on the parameter space
have been placed by requiring the contribution of such particles to dark matter energy
density be in the range allowed by cosmological observations. Search in collider
experiments in LEP 200, LHC and Tevatron is on, but it will be several years before we
have results.}

\end{itemize}
\section{Concluding Remarks}

In this brief review, I have not touched upon a multitude of other ideas and problems,
particularly problems associated with \emph{Dark Energy}. The enormous progress in
observational cosmology and the the unprecedented accuracy of the cosmological
parameters have posed profound problems for both particle physics and cosmology.  It is
clear that the standard model of elementary particles and their interactions fails to
provide a complete catalogue of the building blocks of our Universe. Physics beyond the
Standard Model, Grand Unified Theories and Supersymmetry have hints that they may
provide the necessary ingredients, but it is far from clear. There is also the over
riding problem of baryon asymmetry.The symmetry between particles and antiparticles is
firmly established in collider physics, yet there is no sign of that symmetry in the
observed universe. The observed universe is composed almost entirely of matter with
little or no primordial antimatter. There are various proposals to explain this
asymmetry invoking violation of lepton number (L) during electro/weak phase transition
(Leptogenesis) or the violation of (Baryon number -Lepton Number) during the phase
transition at the grand unification scale (Baryogenesis)\cite{TR}. There is no dearth of
new ideas (Extra Dimensional (large and small)), our universe as a "Brane" in a
multidimensional space-time and so on. The inflationary standard model of cosmology has
many problems of its own when it comes to details. Big questions remain to be answered.
Did inflation occur at all? What is the origin of the hypothetical "inflaton" field that
drove inflation? How did the  different forms of matter/energy of comparable abundance
with transition to accelerated expansion in the present epoch? In any case, the strong
symbiotic relation between particle physics and astrophysics and cosmology has produced
many new challenges.

\center{Acknowledgements}

This work was supported in part by the U.S. Department of energy (DOE) under contract
no. DE-FG02-85ER40237. I am greatly indebted to Mark Trodden for many helpful
discussions.


\begin{thebibliography}{7}

\bibitem{FT}

\emph{Measuring and understanding the Universe},Wendy L. Freedman and Michael S. Turner,
arXiv:astro-ph/0308418

\bibitem{T}

\emph{The New Cosmology}, Michel Turner, arXiv:astro-ph/02022007; \emph{Dark Matter and
Dark Energy; The Critical Questions}, Michel Turner,arXiv:astro-ph/0207297




\bibitem{Feng}

\emph{Supersymmetry and Cosmology}, Jonathan Feng, \emph{Slac Summer Institute, July
28-August 8, 2003, Stanford , California}

\bibitem{TR}

\emph{Recent Progress in Baryogenesis}, Antonio Riotto and Mark Trodden,
Annu.Rev.Nucl.Part.sci.1999.49:35-75


\end{thebibliography}
\end{document}